\documentclass[pra,twocolumn,superscriptaddress,showpacs]{revtex4}

\usepackage{graphicx}%% needed for the figures
\usepackage{subfigure}% needed for the figures in one figure
\usepackage{amsmath,bm,amssymb, bbm}%% to get a few extra environment and symbols

\makeatletter
\newcommand{\Exp}[1]{\,\mathrm{e}^{\mbox{\footnotesize$#1$}}}
\newcommand{\I}{\mathrm{i}}
\newcommand{\tr}{\mathop{\mathrm{Tr}}}
\newcommand{\inn}[2]{\bm{#1}\cdot\!\bm{#2}}

\newcommand{\p}{{\cal P}}
\newcommand{\q}{{\cal Q}}
\newcommand{\vu}{{\cal V}}
\newcommand{\hP}{{\cal H}_{P}}
\begin{document}

\title{Entanglement detection from interference fringes in atom-photon systems}
\author{Jun Suzuki}
%\email{junsuzuki@nii.ac.jp}
\affiliation{National Institute of Informatics, 2-1-2 Hitotsubashi, Chiyoda-ku, Tokyo 101-8430, Japan}
\author{Christian Miniatura}
\affiliation{INLN, UMR 6618, Universit\'e de Nice-Sophia, CNRS; 1361 route des Lucioles, F-06560 Valbonne, France}
\affiliation{Centre for Quantum Technologies, National University of Singapore, 3 Science Drive 2, 117543, Singapore}
\author{Kae Nemoto}
\affiliation{National Institute of Informatics, 2-1-2 Hitotsubashi, Chiyoda-ku, Tokyo 101-8430, Japan}

\date{\today}
%03.65.Wj: State reconstruction, quantum tomography 
%03.67.Mn Entanglement production, characterization and manipulation
%42.25.Hz Interference

\begin{abstract}
A measurement scheme of atomic qubits pinned at given positions is studied by analyzing 
the interference pattern obtained when they emit photons spontaneously. In the case of two qubits, 
a well-known relation is revisited in which the interference visibility is equal to the concurrence 
of the state in the infinite spatial separation limit of the qubits. 
By taking into account the superradiant and subradiant effects, it is shown that 
a state tomography is possible when the qubit spatial separation is comparable to the wavelength of the atomic transition. 
In the case of three qubits, the relations between various entanglement measures 
and the interference visibility are studied, where the visibility is defined from the two-qubit case. 
A qualitative correspondence among these entanglement relations is discussed. 
In particular, it is shown that the interference visibility is directly related 
to the maximal bipartite negativity. 
\end{abstract}

\maketitle
%%%%%%%%%%%%%%%%%%%%%%%%%%%%%%%%%%%%
%%%%%%%%%%%%%%%%%%%%%%%%%%%%%%%%%%%%

\section{Introduction} 

Much attention has been raised recently to the study of atom-photon systems, 
such as ultracold atoms and trapped ions, in view of quantum information 
and computation processing protocols \cite{Rev}. 
Achieving these protocols with hybrid systems requires the capability of manipulating 
atoms and photons with high precision as well as long coherence time for qubits. 
As an example of such demands, an ability to prepare highly entangled states 
is needed in the beginning of protocols. 
It is then necessary to verify the quality of these entangled states and 
this can be accomplished by performing a quantum state tomography \cite{PR}. 
However, it is rather difficult to perform the full tomography, in general, 
and it becomes impractical as the size of Hilbert space increases. 
An alternative to the quantum state tomography is to perform a set of 
measurements to detect the amount of entanglement in the state. It is the main 
objective of this paper to investigate a passive measurement scheme for 
entanglement detection by analyzing the interference pattern of 
photons emitted by atoms. 

Since entanglement and correlation are closely related to each other, a relationship 
between the interference visibility and entanglement in bipartite systems has 
been anticipated in the context of wave-particle duality \cite{JSV,EB}. 
Jacob and Bergou observed that two-qubit concurrence 
is equivalent to two-particle interference visibility in the case of pure states \cite{JB03}. 
To be self-contained, let us consider a simplified argument for the known 
correspondence between the visibility and the concurrence by 
the detection of released photons in the far-field \cite{JSV, JB03}. 
Consider two two-level atoms in which energy levels are $|e\rangle$ and $|g\rangle$, respectively. 
Assume an initial atomic state living in the subspace spanned by 
$|0\rangle=|eg\rangle$ and $|1\rangle=|ge\rangle$, 
\begin{equation} \label{state}
\rho=
\left(\begin{array}{cc}
 |0\rangle,|1\rangle \end{array}\right)
\left(\begin{array}{cc}
\rho_{00}&\rho_{01} \\
\rho_{10}&\rho_{11}
 \end{array}\right)
\left(\begin{array}{c}\langle 0|\\
\langle 1|\end{array}\right)\, , 
\end{equation}
whose concurrence is $C(\rho)=2|\rho_{01}|$. 
The spontaneous emission process of a single photon (i.e., $\rho\to|gg\bm{k}\rangle$) 
occurs with equal probability but different phases. 
Ignoring the details of the coupling between the atoms and the radiation field, 
the standard Michelson's visibility is defined by 
\begin{equation} \label{defvis}
\vu(\rho)=\frac{I_{{\rm max}}-I_{{\rm min}}}{I_{{\rm max}}+I_{{\rm min}}}\, , 
\end{equation}
where $I_{{\rm max}}=\max_{0\le\Phi<2\pi}\tr(I(\Phi)\rho)$ and 
$I_{{\rm min}}=\min_{0\le\Phi<2\pi}\tr(I(\Phi)\rho)$ 
are the maximum and the minimum of interference fringes, respectively, 
and the projection measurement of the phase $\Phi$ is given by 
$I(\Phi)=|\Phi\rangle\langle\Phi|$ with 
$|\Phi\rangle=(|0\rangle+\Exp{\I \Phi}|1\rangle)/\sqrt{2}$. 
It is straightforward to evaluate the visibility in this model as 
\begin{equation} \label{con}
\vu(\rho)=2 |\rho_{01}|=C(\rho)\, , 
\end{equation}
and this completes the claim \cite{comment2}. 
We emphasize that visibility here is not the two-particle visibility defined 
from a combination of correlation functions, which is shown to be equivalent to the concurrence of 
the pure state spanned by $|gg\rangle,|eg\rangle,|ge\rangle,|ee\rangle$ \cite{JSV, JB03}. 

Several generalizations addressing the case of mixed states, of two qudits, and 
of multipartite qubits have also been reported recently \cite{Durr,Tessier, HCO,JB07,EKKC}. 
Despite these progresses, there is still no net conclusive result on general multipartite systems. 
This is simply due to the fact that no generally accepted entanglement measure 
and no unique definition of interference visibility exist in multipartite systems. 
The former, for example, can be seen from the fact that none of these existing measures 
can be used to order the entangled states uniquely \cite{EP, WNGKMV,VADM}. 
The latter problem was addressed in Ref.~\cite{EKKC} where a systematic construction 
for the visibility and the so-called predictability were introduced starting 
from minimal requirements for a proper definition of these concepts. 

In this paper, we analyze the interference pattern generated by photons spontaneously 
emitted by a set of two-level atoms and we investigate the relation between entanglement 
and interference visibility. To explore various kinds of two-qubit and three-qubit entangled states, 
we consider identical two-level atoms pinned at given positions and initially prepared 
in a superposition of the first excited states of the atoms. In the course of time, 
a photon is later released from the excited atom in any direction. In our setting, 
quantum statistical effects play no role and complications related to the atomic motion, 
such as Doppler and recoil effects, are avoided. In an experimental situation, 
this would be realized, for example, with atoms trapped in different potential wells 
in the Lamb-Dicke regime.

Another motivation of this work is to clarify the meaning of the standard interference 
visibility \eqref{defvis} defined in two-path interferometers 
when it is applied to three-path interferometers. In Ref.~\cite{EKKC} for example, 
it was shown that there is an infinite family of state functions which could be 
considered as equally good measures of the interference strength. 
In the present paper, however, rather than studying these many other alternatives, 
the usual standard definition \eqref{defvis} is employed to examine what one can learn from it. 

The content of this paper is as follows. Section II provides a brief summary 
for the known results on two two-level atoms interacting with the radiation field. 
In Sec.~III the interference pattern is analyzed when the emitted photon is detected 
in the far-field and is compared with the entanglement present in the initial 
(possibly mixed) state. It is confirmed that the interference visibility converges 
to the concurrence when the spatial separation of the qubits goes to infinity. 
In Sec.~III C, a state tomography is shown to be possible when the separation 
between the qubits is on the order of the wavelength of the atomic transition. 
In Sec.~IV the previous analysis is extended when the initial atomic qubit state is pre- pared 
in a W-like pure state. The relationship is studied between the interference visibility 
and known entanglement measures. In particular, the interference visibility 
is shown to be able to detect bipartite entanglement. 
A brief discussion is given in Sec.~IV D on a state tomography 
for the W-like state. A summary and possible extensions of the work are stated 
in Sec.~V. The Appendix contains the definitions of the entanglement measures used in the paper.

%%%%%%%%%%%%%%%%%%%%%%%%%%%%%%%%%%%%
%%%%%%%%%%%%%%%%%%%%%%%%%%%%%%%%%%%%
\section{Superradiance and Subradiance from radiative corrections} 

We briefly review some known results 
for two atoms interacting with the quantized radiation field. 
A more detailed account can be found in Ref.~\cite{CT}.

Consider two neutral two-level atoms which are pinned at given positions $\bm{x}_{j}$ ($j=1,2$). 
The internal structure of each isolated atom consists of a unique excited state $|e\rangle$ 
at energy $\hbar\omega_{e}$ with radiative lifetime $\tau = 1/\Gamma$, which is separated by the energy 
$\hbar\omega_0= \hbar c k_0=2\pi\hbar c/\lambda_0$ from a unique ground state $|g\rangle$ 
at energy $\hbar\omega_{g}$. Throughout the paper, a simple scalar model 
for the atom-field interaction is considered and any polarization effects are neglected. 
Fixing the origin of energy at the ground-state levels (i.e., at $2\hbar\omega_{g}$) the free atoms-field 
Hamiltonian $H_0$ for this case reads
\begin{equation}
H_{0}=\sum_{j=1,2} \hbar\omega_0 \ |e_{j}\rangle\langle e_{j}|
+\sum_{\bm{k}}\hbar \omega a^{\dagger}_{\bm{k}}a_{\bm{k}}\, , 
\end{equation}
where $a_{\bm{k}}$ and $a^{\dagger}_{\bm{k}}$ stand for the annihilation 
and creation operators of a photon with momentum $\bm{k}$ and angular frequency 
$ \omega = c |\bm{k}|$. The total Hamiltonian is $H=H_0+V$, 
where the coupling of atoms to the radiation field is given by
\begin{equation}
V=-\sum_{j=1,2}d_{j} E(\bm{x}_{j})\, ,
\end{equation}
in the dipole approximation. Here the scalar radiation field $E(\bm{x})$ reads
\begin{equation}
E(\bm{x})=\I \sum_{\bm{k}}{\cal E}_{\omega}
\left(\Exp{\I \inn{k}{x}}a_{\bm{k}}- \Exp{-\I \inn{k}{x}}a^{\dagger}_{\bm{k}}\right)\, , 
\end{equation}
where ${\cal E}_{\omega}=\sqrt{\hbar\omega/(2\epsilon_{0}L^{3})}$ is the field strength 
at energy $\hbar\omega$ and $L$ is the size of  the quantization box used to define 
the photon modes. 
The dipole operator for the $j$th atom is 
\begin{equation}
d_{j} =d \ (|e_{j}\rangle\langle g_{j}|+|g_{j}\rangle\langle e_{j}|)\, .
\end{equation}
Within the scalar radiation model, the dipole strength $d$ relates to the radiative width of the excited state 
through $\Gamma= d^2k_0^3/(2\pi\hbar\epsilon_0)$. 
In the following, when there is no possible ambiguity for labeling the internal atomic states, 
the atomic index $j$ is omitted. 

Consider the situation where at time $t=0$ the radiation field is 
in its vacuum state $|\bm{0}\rangle$, while one of the atomic qubits is 
in the excited state. In other words the system starts in the subspace $\hP$ 
spanned by the first excited states without photons 
(i.e., $|eg\bm{0}\rangle$ and $|ge\bm{0}\rangle$). 
Because these first excited states are coupled to the radiation field, 
they eventually decay to the atomic ground state $|gg\rangle$ 
by releasing a spontaneous photon $|\bm{k}\rangle$.  
This process can be described by the method of the resolvent operator $G(z) = (z-H)^{-1}$ 
and its projection on the various subspaces of interest \cite{CT}. 
For example, the time-evolution operator restricted to the subspace $\hP$ 
is obtained from a contour integral of the following projected resolvent operator 
\begin{align}
&G_{P}(z)=\p\frac{1}{z-H_{0}-V}\p\, ,\\ 
&\p=|eg\bm{0}\rangle\langle eg\bm{0}|+|ge\bm{0}\rangle\langle ge\bm{0}|\, ,
\end{align}
where $\p$ is the projector onto the subspace $\hP$. 
To second order in the coupling constant, its diagonalized form reads
\begin{equation}
G_{P}(z) \simeq |+\rangle\frac{1}{z-\hbar\Omega_{+}} \langle +|+
|-\rangle\frac{1}{z-\hbar\Omega_{-}} \langle -|\, ,
\end{equation}
where the corresponding eigenkets $|\pm\rangle$ are given by 
\begin{equation} \label{eigenpm}
|\pm\rangle=\frac{1}{\sqrt{2}}(|eg\bm{0}\rangle\pm|ge\bm{0}\rangle)\, , 
\end{equation}
and the complex eigenvalues $\Omega_{\pm} = \omega_\pm - \I  \Gamma_\pm/2$ are \cite{comment5}
\begin{align}
&\omega_{\pm}=\omega_0\mp  \frac{\Gamma}{2} g(k_0r)\, ,\\
&\Gamma_\pm= \Gamma\left[1\pm f(k_0r)\right]\, , \\  \label{2-RadStates}
&g(x)=\cos(x)/x, \ f(x)=\sin(x)/x\, ,
\end{align}
where $r=|\bm{x}_1-\bm{x}_2|$ is the distance between the atoms.

The net effect of the interaction with the radiation field is to lift the degeneracy 
between the initial atomic states in the subspace $\hP$. The physical process 
behind it is the resonant exchange of photons between the atoms which ``glues'' the atoms 
together and gives rise to the states $|\pm\rangle$ in a fashion similar to the bonding 
and antibonding states of a molecule. The new levels also acquire different finite lifetimes. 
The subspace $\hP$ is thus irreversibly emptied by spontaneous emission 
in the course of time while the ground state is gradually populated. 
Both the degeneracy splitting and the lifetimes depend on the relative distance $r$ between the atoms. 
For sufficiently close atoms, the celebrated superradiant and subradiant behaviors 
are recovered in which the eigenstates are given by $|+\rangle$ and $|-\rangle$, respectively. 
The atomic dipoles are perfectly correlated (i.e., they oscillate in phase for the superradiant state 
and in phase opposition for the subradiant state). For $|+\rangle$, the radiation waves emitted 
by the two atoms interfere constructively and the system radiates more efficiently, shortening its lifetime. 
For $|-\rangle$, on the other hand, the waves interfere destructively and the system cannot radiate, 
increasing its lifetime. Indeed one gets $\Gamma_{+} \to 2\Gamma$ 
and $\Gamma_{-} \to 0$ in the limit $k_0r \to 0$ \cite{CT}. 

In the other extreme limit $k_0r\to\infty$, both lifetimes achieve the value 
obtained for a single isolated atom $\Gamma_{\pm}\to\Gamma$. 
In this limit the atoms are no longer coupled by the radiation field, 
and the energy degeneracy in the subspace $\hP$ shows up again 
$\omega_{\pm}\to\omega_0$. 

The time evolution of the initial state decaying into the ground state is obtained 
from a contour integral of another projected resolvent 
\begin{equation}
G_{QP}(z) =\q\frac{1}{z-H_{0}-V}\p,\quad \q=1-\p\, . 
\end{equation}
The matrix elements of $G_{QP}(z)$ read 
\begin{equation}  \label{result}
\langle gg\bm{k}| G_{QP}(z)|\pm\rangle 
\simeq\frac{\I \hbar g_{\omega}}{\sqrt{2}}\ \frac{\Exp{-\I \inn{k}{x}_1}\pm\Exp{-\I \inn{k}{x}_2}}
{(z-\hbar\omega)(z-\hbar\Omega_{\pm})}\, ,
\end{equation}
where $g_{\omega}=d{\cal E}_{\omega}/\hbar$ is the Rabi frequency at field angular frequency 
$\omega$. One can easily check that $\langle gg\bm{k}| G_{QP}(z)|-\rangle \to 0$ 
when $k_0r \to 0$ as expected for the subradiant state 
since it does not couple anymore to the radiation field in this limit. 

%%%%%%%%%%%%%%%%%%%%%%%%%%%%%%%%%%%%
%%%%%%%%%%%%%%%%%%%%%%%%%%%%%%%%%%%%
\section{Entanglement detection: The case of two atoms} 

%%%%%%%%%%%%%%%%%%%%%%%%%%%%%%
\subsection{Visibility in the infinite separation limit}

Let us now proceed to verify the result (\ref{con}) based on the microscopic model presented 
in the previous section. As before, consider the initial state (\ref{state}) 
of atoms within the subspace spanned by $|ge\bm{0}\rangle$ and $|eg\bm{0}\rangle$, 
and parametrize the density matrix as 
\begin{equation} \label{2state}
\rho_{2}=
(|eg\bm{0}\rangle,\ |ge\bm{0}\rangle)
\frac{1}{2}
\left(\begin{array}{cc}
1\!+\!s_{z}&s_{x}\!-\!\I s_{y} \\
s_{x}\!+\!\I s_{y}& 1\!-\!s_{z}
\end{array}\right)
\binom{\langle eg\bm{0}|}{
\langle ge\bm{0}|}\, ,
\end{equation}
where $s_x=s\sin\theta\cos\phi$, $s_y=s\sin\theta\sin\phi$, 
and $s_z=s\cos\theta$ are the components of the vector $\bm{s}$ with 
the ranges $0\le\theta\le\pi$, $0\le\phi\le2\pi$, and $0\le s\le 1$. 
The concurrence of the state \eqref{2state} is $C(\rho_{2})=s \sin\theta$. 

In the model considered, the interference visibility (\ref{defvis}) is defined by the maximum 
and the minimum values achieved by the spectral distribution of the spontaneous photon 
over all possible emission directions. This spectral distribution is proportional to the transition probability 
$P(\rho_{2}\to |gg\bm{k}\rangle)$ from the initial state (\ref{2state}) to the ground state $|gg\bm{k}\rangle$. 
This is obtained from the time evolution of the state $\rho_{2}(t)=U(t)\rho_{2}U^{\dagger}(t)$ in the long-time limit 
$\Gamma t\to\infty$ where $U(t)$ is the time-evolution unitary operator calculated from (\ref{result}) as follows: 
\begin{align} \label{dist}
&P(\rho_{2}\to |gg\bm{k}\rangle)=\lim_{\Gamma t \to \infty}
\tr\left(|gg\bm{k}\rangle\langle gg\bm{k}|\rho_{2}(t) \right)\\ \nonumber
&=\!\frac{ \Gamma\omega}{(2\pi)^{2}k_{0}^3}
\left[ \frac{{\cal B}_{+}}{(\omega\!-\!\omega_{+})^{2}\!+\!(\Gamma_{+}/2)^{2}}\!+\!
\frac{{\cal B}_{-}}{(\omega\!-\!\omega_{-})^{2}\!+\!(\Gamma_{-}/2)^{2}} \right] ,  
\end{align}
where the coefficients ${\cal B}_{\pm}$ are 
\begin{equation} \label{DefB} \nonumber
{\cal B}_{\pm}\!=\!\frac{1}{4}\left[(1\pm s_{x})(1\pm\cos\bm{k}\!\cdot\!\bm{r})
\!+\!\frac{1\!\pm\! f}{1\!+\!g^{2}}(s_{y}-gs_{z} )\sin\bm{k}\!\cdot\!\bm{r}  \right],  
\end{equation}
with the shorthands $f \equiv f(k_0r)$ and $g \equiv g(k_0r)$, 
and $\bm{r}=\bm{x}_{1}-\bm{x}_{2}$ is the relative vector connecting the two atoms. 
In Eq.~\eqref{dist}, the spectral distribution is a weighted sum of two Lorentzians centered at 
angular frequencies $\omega_{\pm}$ with respective widths $\Gamma_{\pm}$. 
The angular frequency separation $\Delta\omega = |\omega_{+}-\omega_{-}|$ 
between the two peaks is generally small and the atoms need to be located rather close to 
each other to distinguish them. This can be seen from the ratio 
\begin{equation}
\frac{\Delta\omega}{\Gamma_{\pm}}=\left|\frac{g(k_{0}r)}{1\pm f(k_{0}r)}\right|\, , 
\end{equation}
which is a small number when atoms are located far apart compared with the optical atomic transition wavelength. 

To extract the visibility of the interference pattern, 
the distribution (\ref{dist}) is to be maximized and minimized 
over the spherical angles of the emitted photon. Rewrite (\ref{dist}) as
\begin{equation}
P(\rho_{2}\!\to\!|gg\bm{k}\rangle)\!=\!\frac{ \Gamma\omega}{(2\pi)^{2}k_{0}^3}\left[\xi_{+}\!+\!\sqrt{\xi_{-}^{2}\!+\!\eta^{2}}
\cos(\bm{k}\!\cdot\!\bm{r}-\theta_{0}) \right] , 
\end{equation}
with the notations 
\begin{align}\label{xi}
&\xi_{\pm}\!=\!\frac{1+s_{x}}{(\omega\!-\!\omega_{+})^{2}\!+\!(\Gamma_{+}/2)^{2}}\pm
\frac{1-s_{x}}{(\omega\!-\!\omega_{-})^{2}\!+\!(\Gamma_{-}/2)^{2}}\, ,\\ \label{eta} \nonumber
&\eta\!=\!\left[ \frac{1+f}{(\omega\!-\!\omega_{+})^{2}\!+\!(\Gamma_{+}/2)^{2}}+
\frac{1-f}{(\omega\!-\!\omega_{-})^{2}\!+\!(\Gamma_{-}/2)^{2}}\right]\\
&\hspace{5.5cm}\times\frac{s_{y}-gs_{z} }{1+g^{2}}\, ,
\end{align}
and $\theta_{0}=\tan^{-1}(\eta/\xi_{-})$. It is clear that the maxima and minima 
occur when $\cos(\bm{k}\!\cdot\!\bm{r}-\theta_{0})=\pm1$.
The far-field fringe visibility is thus 
\begin{equation} \label{vis2}
\vu(r, \omega ;\rho_{2})=\frac{\sqrt{\xi_{-}^{2}+\eta^{2}}}{\xi_{+}}\, .
\end{equation}
It is straightforward to calculate the visibility in the infinite separation limit as 
\begin{equation} \label{inflim}
\lim_{r\to\infty}\vu(\omega, r;\rho_{2})=\sqrt{s_{x}^{2}+s_{y}^{2}}=s \sin\theta=C(\rho_{2})\, . 
\end{equation}
Therefore, the result \eqref{con} is valid only in the infinite separation limit. 
From a practical point of view, the infinite separation limit is reached as soon as 
the two atoms are separated by a large distance compared to the wavelength 
$\lambda_0$ of the optical transition of each individual atom. 

A full characterization of the initial state $\rho_2$ is completed 
as soon as the full vector $\bm{s}$ associated to $\rho_2$ is known, see (\ref{2state}). 
Note that the angle $\theta_{0}$ describes a shift of the interference pattern 
perpendicular to the plane located at equal distances from the atoms and that 
it converges to the angle $\phi$ in the limit $r\to\infty$. Thus, knowing the positions $\bm{x}_j$ 
of the atoms, the pattern shift can be measured in principle and $\phi$ can be extracted from the data. 
From the visibility and the interference pattern shift, $s_x$ and $s_y$ can be obtained. 
However one still needs to extract the missing component $s_z$ from the data, 
which is unfortunately not possible in the infinite separation limit. 
Complete state tomography is studied in Sec.~III C. As a last remark, 
the angular separation between two consecutive fringes is inversely proportional to distance $r$. 
This means that the further apart atoms are located, 
the more periodic and regular the fringe pattern appears. 
Therefore, by knowing the distance between the two atoms, one does not need to measure 
the whole interference pattern for all angular positions. It is sufficient to record a few fringes 
to detect the amount of entanglement. 

%%%%%%%%%%%%%%%%%%%%%%%%%%%%%%%%%%%%
\subsection{Interference visibility for finite distances}

In this section, we analyze the deviation $\vu(r,\omega; \rho_2)-C(\rho_2)$ of the interference visibility 
obtained for atoms at finite distances, Eq.~(\ref{vis2}), from its asymptotic value obtained for atoms far apart, 
that is, from the concurrence $C(\rho_2)=s\sin\theta$ of the two-qubit initial state.
Our interest is in the state and distance dependency of this deviation 
and particularly in the maximal possible deviation from the concurrence 
for a given distance $r$ between the atoms. The latter number can be used as a quantitative 
measure for the relationship between interference visibility and entanglement 
when estimating the amount of entanglement in the initial state from the observed visibility. 

\begin{figure}[htbp]
\begin{center}
\includegraphics[width=3.3in,keepaspectratio,clip]{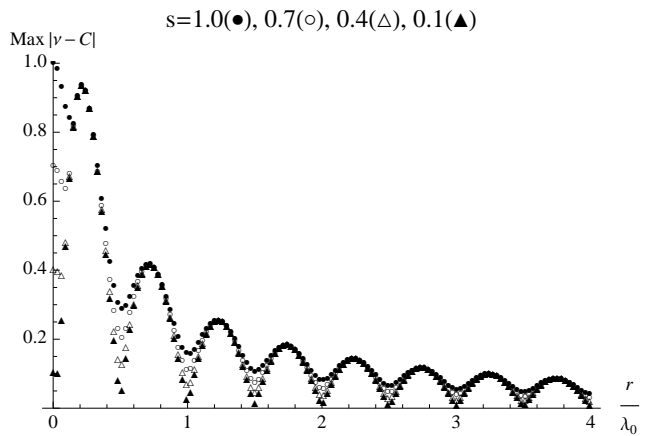}
\caption{Plot of $\max_{\theta,\phi}|\vu-C|$ as a function of $r$ (in units of $\lambda_0$) 
for different values of the purity $s=1.0, 0.5, 0.1$. The photon angular frequency has been set at 
$\omega_0$. The local minima occur when $f(k_0r)=0$ (i.e., when $r$ is a multiple integer of $\lambda_{0}/2$) 
while the local maxima occur when $g(k_0r)=0$ (i.e., when $r$ is an odd multiple integer of $\lambda_{0}/4$). 
The values achieved by the local maxima are purity independent while the values achieved 
by the local minima increase with the purity. The solid line represents the analytical result for $s=0$, Eq.~\eqref{s0decay}. 
}
\label{fig1}
\end{center}
\end{figure}

First, from the result (\ref{vis2}) it can be seen that the deviation can be both positive and negative, 
and converges to zero in the limit $r\to\infty$ as was shown in (\ref{inflim}). 
Next, we numerically calculate the maximal value reached by $|\vu-C|$ as a function of $r$ 
(in units of $\lambda_0$) for some specific values of the state purity $s=1.0, 0.5, 0.1$. 
To this end, the deviation $|\vu-C|$ is maximized over the state parameters $\theta$, $\phi$ at fixed $r$ and $s$. 
The frequency of the emitted photon is set at $\omega_0$. This can be achieved, for example, 
by frequency filtering in the detection process. Without such filtering, one would have to integrate over all range of frequencies 
in Eq.~\eqref{dist} to get the observed fringes. 
The result, depicted in Fig.~\ref{fig1}, shows an oscillatory decay of 
$\max_{\theta,\phi}|\vu-C|$ when $r$ is increased. The local minima occur 
when $k_0r$ is an integer multiple of $\pi$, that is, when $f(k_0r)=0$. 
For these particular atomic separations, the superradiant and subradiant states have 
exactly the same decay rate $\Gamma_{\pm}=\Gamma$ 
but still different eigenfrequencies. The local maxima occur 
when $k_0r\pm\pi/2$ is a multiple integer of $2\pi$, that is, 
when $g(k_0r)=0$. In this case, the superradiant and subradiant states have 
identical eigenfrequencies $\omega_{\pm}=\omega_0$, but achieve different decay rates.
As one can see, the values of the local maxima are insensitive to the purity of the state 
whereas the values of the local minima increase monotonically when the purity of the state is increased. 

For $s=0$, $C(\rho_2)=0$ and we get the analytical result 
\begin{equation}\label{s0decay}
|\vu-C|=\frac{2 k_0r |\sin (k_0r)|}{1+(k_0r)^2}. 
\end{equation}
It may seem counterintuitive that one gets nonzero fringe visibility 
when the purity of the initial state is $s=0$ (i.e., when the concurrence of the state is zero 
and no coherence seems present in the system). In this case, the initial state $\rho_2$ describes 
a complete statistical mixture of the qubit states $|eg\rangle$ and $|ge\rangle$ 
and one can hardly imagine measuring interference fringes using 
a Young slit device operating with one slit shut. 
The reason is that, due to the resonant exchange of photons 
between the atoms, the radiating eigenstates are {\it not} $|eg\bm{0}\rangle$ 
and $|ge\bm{0}\rangle$ but the superradiant and subradiant states 
$|\pm\rangle=(|eg\bm{0}\rangle\pm |ge\bm{0}\rangle)/\sqrt{2}$. 
The initial state at $t=0$ is also equivalent to a complete statistical mixture of 
$|\pm\rangle$, and both the superradiant and subradiant states 
display well-defined relative phases 
between the states $|eg\rangle$ and $|ge\rangle$. Both states $|\pm\rangle$ 
lead to an interference pattern independently in the far-field, 
the dark fringes of one pattern corresponding to the bright fringes of the other and vice verse. 
In addition, these states decay with different rates. 
Therefore the sum of their respective interference patterns does not average out except 
in the limit $r\to\infty$. This is easily seen from Eqs.~(\ref{xi}) and (\ref{eta}) where $\eta=0$ and 
$\theta_0=0$ for $s=0$ and hence the visibility of 
the interference pattern for $s=0$ relates to the difference of the Lorentzians. 
For atoms that are far apart, the Lorentzians are essentially identical and cancel out, and the interference is lost. 
In this limit, the intuitive result is recovered in which the radiation takes place starting from 
the complete statistical mixture of the qubit states $|eg\rangle$ and $|ge\rangle$. 
For atoms that are close enough, however interference fringes do occur as the result of the overlap 
and partial blurring of two independent interference patterns, associated to 
$|+\rangle$ and $|-\rangle$, with different strengths and fringe locations.

As can be seen from Fig.~\ref{fig1} and Eq.~\eqref{s0decay}, large deviations 
between $\vu(r,\omega;\rho_2)$ and $C(\rho_2)$ appear 
for small separation of the atoms, typically in the region $r<\lambda_{0}$. However, 
the deviation can still be significant for atoms far apart as it only decays like $2/k_0r$ for large spatial separations. 
It is straightforward to understand why the argument presented in the Introduction 
to derive the relation between the visibility 
and concurrence applies to atoms that are far apart. Concurrence relates to entanglement present 
in the atomic state $\rho_2$ spanned by the states $|eg\rangle$, $|ge\rangle$ irrespective 
of the coupling to the radiation field while visibility refers to the interference pattern 
obtained by coupling the atoms to the radiation field 
(i.e., it relates to the interference pattern obtained from the radiating states $|\pm\rangle$). 
However, for atoms that are far apart, the resonant exchange of photons between the atoms is negligible 
and the interaction with the vacuum field does not lift the degeneracy of the subspace $\hP$. 
As a consequence, the eigenstructures to analyze radiation and entanglement are then essentially the same. 
This is why, for atoms that are far apart, the entanglement present in the initial state 
becomes equivalent to the visibility of the fringes. 

%%%%%%%%%%%%%%%%%%%%%%%%%%%%%%%%%%%%
\subsection{State tomography}

In this section, it is shown that, for finite atomic separations, the transition probability 
$P(\rho_2 \to |gg\bm{k}\rangle)$ and the visibility $\vu(r,\omega;\rho_2)$ encode 
enough information to reconstruct the full initial state (\ref{2state}). This is obvious from 
Eqs.~(\ref{dist}) and (\ref{vis2}) which can be directly used to 
extract the values $s$, $\theta$, and $\phi$ as soon as $r$ is known. 
Complete knowledge about the initial state then provides information on 
the amount of entanglement in it. 

There are many ways to demonstrate state tomography. 
For example, consider the simple case where $k_{0}r >1$, but is not too large. 
The frequency of the emitted photon is set at $\omega_0$ as before. 
Expanding the transition probability and 
the phase-shift angle up to the first order in $f$ and $g$ reads 
\begin{align} \label{apppro} \nonumber
&P(\rho_{2}\to |gg\bm{k}\rangle)\simeq\frac{8 c}{(2\pi k_{0})^{2} \Gamma} \ [1\!-\!2fs_x\\
&\hspace{1.5cm}+(s_x\!-\!2f)\cos\bm{k}\!\cdot\!\bm{r}
+(s_y\!-\!gs_z)\sin\bm{k}\!\cdot\!\bm{r} ] ,\\
&\theta_{0}\simeq \tan^{-1}\left( \frac{s_y-gs_z}{s_x}-\frac{2fs_y}{s_x^{2}}\right)\, .  \label{appphs}
\end{align}
First of all, knowing $k_{0}r$ calculates the values $f$ and $g$. 
Second, from the probabilities at $\bm{k}\!\cdot\!\bm{r}=0$ or $\pi$ in Eq.~\eqref{apppro}, 
the state parameter $s_x$ is obtained. Similarly, knowing $s_x$ and using the values 
at $\bm{k}\!\cdot\!\bm{r}=\pm\pi/2$, the combination of 
state parameters $s_y-gs_z$ can be found. 
Last, measuring the approximated phase shift $\theta_0$ in 
Eq.~\eqref{appphs} together with the obtained values of $s_x$ and $s_y-gs_z$ provides $s_y$ 
and thus three state parameters $s_x$, $s_y$, and $s_z$. 
Therefore the full vector $\bm{s}$ is known, completing the state tomography. 

The main result of this discussion is that spontaneously emitted photons 
from the excited states (\ref{2state}) do provide full information about the initial state 
as well as the amount of entanglement in it. The drawback of this tomographic scheme is 
that it is not very efficient. Indeed, in the absence of any prior information on the initial state 
that could help simplify the state search, one needs to measure photons emitted 
in all directions to obtain the full fringe pattern and later characterize the state. 
This requires the repeated measurement of photons from an identically prepared state. 
Although the method is not very practical, it can be combined with other proposals.
For example, other possible entanglement detection schemes were studied by many authors. 
Among them, we mention Ref.~\cite{SMM}, where an entanglement witness was constructed 
without using the full interference fringe pattern.
 
%%%%%%%%%%%%%%%%%%%%%%%%%%%%%%%%%%%%
%%%%%%%%%%%%%%%%%%%%%%%%%%%%%%%%%%%%
\section{Interference visibility in the three-atom system}

In this section, we extend the previous analysis to the case of three atoms and 
investigate the relationship between the visibility 
and the various entanglement measures. This can be accomplished 
along the same lines of the microscopic calculation done for the two-atom system. 
As shown in the following, however, a complete analysis appears difficult. 
Two major obstacles are shown in the following. 
The full diagonalization of the resolvent operator does not take a simple 
analytical form except for a few limited cases, for example, when the three atoms are 
pinned at the vertices of an equilateral triangle. 
Even for such a special case, finding the extrema of 
the spectral distribution of photons is unlikely to be carried out by hand, 
and one needs to rely on numerical calculations in general. It is shown 
in the following that the large separation limit reduces 
the optimization problem significantly. 
Therefore, we shall investigate the relations between 
interference visibility and various known entanglement measures for 
three-atom systems in this large separation limit. The full analysis 
including resonant atom-photon interactions is to be studied in a future publication. 

Consider the simple geometry where the atoms are pinned 
at the vertices of an equilateral triangle with equal mutual distance $r$. 
Take their positions at $\bm{x}_j=r(\sin\vartheta_{j},-\cos\vartheta_{j},0)/\sqrt{3}$ 
with the angle $\vartheta_{j}=2\pi(j-1)/3$ ($j=1,2,3$). As for the two-atom case, 
the initial atomic system is the subspace spanned by 
$|egg\rangle$, $|geg\rangle$, and $|gge\rangle$ 
while the radiation field starts in its vacuum state $|\bm{0}\rangle$. 
This subspace of the initial state is referred to as the subspace $\hP$ as before. 
The visibility of the far-field interference pattern is calculated when a photon 
is spontaneously released in the mode $|\bm{k}\rangle$ with angular frequency $\omega=c|\bm{k}|$ 
and the atomic system ends in its ground state $|ggg\rangle$. 
Moreover, instead of considering a general initial mixed atomic state 
as was done in the previous sections, 
the following family of pure states, known as the W-like states, is investigated 
\begin{equation}  \label{3state}
|\psi\rangle=c_{1}|egg\rangle+c_{2}|geg\rangle+c_{3}|gge\rangle\, ,
\end{equation}
with the normalization $c_{1}^{2}+c_{2}^{2}+c_{3}^{2}=1$. 
Without loss of generality, the coefficients $c_{j}$ can be set as positive numbers 
fulfilling $c_{1}\ge c_{2}\ge c_{3}>0$. The conditions $c_{j}>0$ are imposed, 
otherwise the problem reduces to the two-atom case when one of them is equal to zero. 
Such a choice is always possible since any state with arbitrary complex coefficients 
$c_{j} \exp(\I \phi_{j})$ can always be unitarily mapped onto the state 
\eqref{3state} by applying the following tensor product of local phase gates, 
\begin{equation} \nonumber
\left(\begin{array}{cc}1 & 0 \\0 & \Exp{-\I\phi_{1}}\end{array}\right)\!\otimes\!
\left(\begin{array}{cc}1 & 0 \\0 & \Exp{-\I\phi_{2}}\end{array}\right)\!\otimes\!
\left(\begin{array}{cc}1 & 0 \\0 & \Exp{-\I\phi_{3}}\end{array}\right)\, .
\end{equation}
According to the stochastic local operation and classical communication classification of 
entanglement, the state Eq. \eqref{3state} belongs to the W-state class and as such its tangle 
is equal to zero \cite{DVC}. However, we comment that, even for the pure three-qubit system, 
there is no generally accepted entanglement measure which is able to fully and 
uniquely characterize the amount of entanglement.

%%%%%%%%%%%%%%%%%%%%%%%%%%%%%%%%%%%%
\subsection{Superradiance and subradiance effects for the three-atom case}

Similar to the case of two atoms, we briefly state the results for the interaction 
of the three-atom system, initially prepared in the subspace $\hP$, 
with the scalar radiation field initially starting in its vacuum state $|\bm{0}\rangle$. 
As a result of the resonant exchange of photons between the atoms, 
the energy degeneracy of this subspace is split in two, 
one superradiant state and two degenerate subradiant states. 
The projected resolvent operator describing the evolution of the system 
in the subspace $\hP$ is given as 
\begin{multline}
G_{P}(z) \simeq |+\rangle\frac{1}{z-\hbar\Omega_{+}}\langle+|\\
+|1-\rangle\frac{1}{z-\hbar\Omega_{-}}\langle1-|
+|2-\rangle\frac{1}{z-\hbar\Omega_{-}}\langle2-|\, , 
\end{multline}
where the superradiant and the subradiant eigenstates ($\ell=1,2$) are given by 
\begin{align} 
|+\rangle&=\left(|egg\bm{0}\rangle+|geg\bm{0}\rangle+|gge\bm{0}\rangle\right)/\sqrt{3}\, ,\\
|\ell-\rangle&=(q^{\ell}|egg\bm{0}\rangle+q^{2\ell}|geg\bm{0}\rangle+|gge\bm{0}\rangle)/\sqrt{3}\, , 
\end{align}
where $q=\exp(2\pi\I/3)$. Taking the origin of energies 
at $3\hbar\omega_{g}$ and absorbing the Lamb-shift correction into a re-definition of 
the individual atomic angular transition frequency $\omega_{0}$, 
the superradiant and subradiant eigenvalues 
$\Omega_{\pm} = \omega_{\pm} -\I\Gamma_{\pm}/2$ read 
\begin{align}
\omega_{\pm}&=\omega_{0}-\Gamma \,\frac{3\pm 1}{4}\, g\, ,\\
\Gamma_{\pm}&= \Gamma \left(1\pm\frac{3\pm 1}{2}f\right)\, ,
\end{align}
where $f\equiv f(k_0r)$ and $g\equiv g(k_0r)$ are given in \eqref{2-RadStates}. 
As a result, the superradiant state $|+\rangle$ decays three times faster 
than the isolated atoms and the degenerate subradiant states $|\ell-\rangle$ 
does not decay in the limit $r\to 0$. 
For atoms far apart $r\to\infty$, the limits $\Gamma_{\pm} \to \Gamma$ 
and $\omega_{\pm}\to\omega_0$ hold, and we recover the case of isolated atoms. 

The initial state $|\psi\rangle|\bm{0}\rangle$ living in the subspace $\hP$ 
decays to the atomic ground state $|ggg\rangle$ 
with a spontaneously emitted photon $|\bm{k}\rangle$. 
In a way similar to the two-atom case, the probability of such a process reads 
\begin{widetext}
\begin{equation} \label{3p}
P(|\psi\rangle|\bm{0}\rangle\to|ggg\bm{k}\rangle)\simeq 
\frac{ \Gamma\omega}{(2\pi)^{2}k_{0}^3}
\left[ \frac{{\cal D}_{+}}{(\omega\!-\!\omega_{+})^{2}\!+\!(\Gamma_{+}/2)^{2}}
\!+\!\frac{{\cal D}_{-}}{(\omega\!-\!\omega_{-})^{2}\!+\!(\Gamma_{-}/2)^{2}} \right]\, . 
\end{equation}
The weights of the Lorentzians read 
\begin{align}
{\cal D}_{+}&=\frac{3}{2}\bar{c}^{2}+\sum^{3}_{i>j=1}
\left\{\bar{c}[\bar{c}+(c_{i}+c_{j}-2\bar{c})h_{+}]\cos \bm{k}\!\cdot\!(\bm{x}_{i}-\bm{x}_{j}) 
+\bar{c}(c_{i}-c_{j})h_{0}\sin\bm{k}\!\cdot\!(\bm{x}_{i}-\bm{x}_{j})\right\}\, ,  \\
{\cal D}_{-}&=\frac12 (1-3\bar{c}^{2})+\sum^{3}_{i>j=1}
\left\{[(c_{i}-\bar{c})(c_{j}-\bar{c})+\bar{c}(c_{i}+c_{j}-2\bar{c})h_{-}]\cos\bm{k}\!\cdot\!(\bm{x}_{i}-\bm{x}_{j}) 
+\bar{c} (c_{i}-c_{j})h_{0}\sin\bm{k}\!\cdot\!(\bm{x}_{i}-\bm{x}_{j}) \right\}\, ,
\end{align}
\end{widetext}
where $\bar{c}=(c_{1}+c_{2}+c_{3})/3$, and the coefficients $h_{\pm,0}$ are 
\begin{align}
h_{\pm}&=\frac{2+f}{(3g)^{2}+(2+f)^{2}}\left(1\pm\frac{3\pm1}{2}f \right)\, ,\\
h_{0}&=\frac{3g}{(3g)^{2}+(2+f)^{2}}\, .
\end{align}

%%%%%%%%%%%%%%%%%%
\subsection{Visibility in the large separation limit}

The previous expression for the transition probability (\ref{3p}) is quite complex 
and the large separation limit $k_{0}r\gg 1$ is considered to simplify the results 
and the discussion. In this case, the simplified result reads 
\begin{multline} \label{3plim}
P(|\psi\rangle|\bm{0}\rangle\to|ggg\bm{k}\rangle)\simeq 
\frac{ \Gamma\omega}{2(2\pi)^{2}k_{0}^3}\frac{1}{(\omega-\omega_0)^{2}+(\Gamma/2)^{2}}\\
\times\left[ 1+2\sum^{3}_{i>j=1}c_{i}c_{j}\cos\bm{k}\!\cdot\!(\bm{x}_{i}-\bm{x}_{j}) \right]\, . 
\end{multline}
Indeed, as already discussed in the two-atom case, in the large separation limit 
the radiative decay from state $|\psi\rangle$ is equivalent 
to a triple slit experiment where the photon could have been released with equal probability 
from any of the states $|egg\rangle$, $|geg\rangle$, and $|gge\rangle$. This is because 
the resonant exchange of photons between the atoms is negligible and the degeneracy 
in the subspace $\hP$ is not lifted. The superradiant and subradiant states are 
degenerate in this limit and their radiation properties are identical. 
Note that the coefficient $2c_{i}c_{j}$ is the concurrence 
of the reduced state $\rho_{ij}=\tr_{k\not=i,j}|\psi\rangle\langle\psi|$ obtained 
when the remaining atom has been traced out. 

To extract the visibility of the fringe pattern, the transition probability (\ref{3plim}) 
needs to be maximized and minimized over the emission angles of 
the released photon. This is equivalent to finding the maximum and minimum values of the function 
\begin{equation}
I(\theta_{j})=1+2\left(c_{2}c_{3}\cos\theta_{1}+c_{3}c_{1}\cos\theta_{2}+c_{1}c_{2}\cos\theta_{3} \right)\, , 
\end{equation}
over the angles $\theta_{i}\equiv\bm{k}\!\cdot\!(\bm{x}_{j}-\bm{x}_{k})=
\bm{k}\!\cdot\!\bm{r}_{i}$ with $\bm{r}_{j}=r(\cos\vartheta_{j},\sin\vartheta_{j},0)$. 
Here $(ijk)$ is a cyclic permutation of $(123)$. By definition, 
these angles are linearly dependent as constrained by $\theta_{1}+\theta_{2}+\theta_{3}=0$. 
With this function $I(\theta_j)$, the visibility is expressed as 
\begin{equation}
\vu(|\psi\rangle)=\frac{I_{{\rm max}}-I_{{\rm min}}}{I_{{\rm max}}+I_{{\rm min}}}\, . 
\end{equation}

The maximum is easily found when all cosine functions reach 
the maximum by $\cos\theta_{j}=1$ ($j=1,2,3$), leading to
\begin{equation}\label{Imax}
I_{{\rm max}}=(c_{1}+c_{2}+c_{3})^{2}\, .
\end{equation}
For example a bright fringe is obtained when the photon is emitted 
perpendicularly to the plane of the equilateral triangle made by the three atoms, 
in which case $\bm{k}\!\cdot\!(\bm{x}_{i}-\bm{x}_{j})=0$.

Finding the minimum requires a little bit of caution as there are two cases. 
In the first case, $I_{{\rm min}}=0$, leading to full fringe visibility $\vu=1$. 
As $c_{1}\ge c_{2} \ge c_{3}$, this is the case if and only if the triangle inequality 
$c_{1} \le c_{2}+c_{3}$ is fulfilled. 
This is achieved by the angles satisfying 
\begin{equation}\label{anglemin}
\cos\theta_{j}=\frac{2c_{j}^{2}-1}{2c_{1}c_{2}c_{3}}c_{j}\quad (j=1,2,3) \, .
\end{equation}
On the other hand, if $c_{1}>c_{2}+c_{3}$, the minimum is attained when 
$\cos\theta_{1}=1$ and $\cos\theta_{2}=\cos\theta_{3}=-1$, leading to 
\begin{equation}\label{Imin}
I_{{\rm min}}=(c_{1}-c_{2}-c_{3})^{2}\, . 
\end{equation}
The visibility in this case is less than unity.

Summarizing the previous results, the visibility reads
\begin{equation} \label{3vis}
\vu (|\psi\rangle)=
\begin{cases} 
\quad 1 \quad \hspace{1.55cm}(c_{1}\le c_{2}+c_{3})\\
\frac{2c_{1}(c_{2}+c_{3})}{1+2c_{2}c_{3}}<1\quad (c_{1}> c_{2}+c_{3})\, .
 \end{cases}
\end{equation}

We remark that the so-called W-state $|W\rangle=(|egg\rangle+|geg\rangle+|gge\rangle)/\sqrt{3}$ 
achieves $\vu=1$ as naively expected. 
This is easy to understand in the Young slit language as the triple slits radiate 
in this condition on equal footing. However it is stressed that there are infinitely many other states 
reaching full fringe visibility. Finally, note that the other extreme limiting case $c_{2},c_{3} \to 0$ 
means that only one slit is operating and thus one obtains $\vu=0$ in this case 
at least in the large limit. In the next section, these results are examined 
in the light of known entanglement measures. 

%%%%%%%%%%%%%%%%%%%%%%%%%%%%%%%%
\subsection{Relation between interference visibility and entanglement measures}

Bearing in mind the results of Sec.~III, it is natural to expect that 
the previous three-atom interference visibility (\ref{3vis}) is somehow 
related to the amount of entanglement present in the initial atomic W-like state 
\eqref{3state}. Unfortunately, it is seen that W-like states with $c_{1}\le c_{2}+c_{3}$ 
cannot be discriminated at all since they all lead to $\vu =1$. This raises 
the question of finding the range of entanglement measures compatible 
with a given observed value of the interference visibility. 
Generally speaking, one can infer the amount of entanglement in the state 
with better precision when 
the range of corresponding entanglement measure is smaller. 
Depending upon the protocols used, one may need to know the 
least amount of entanglement to make sure that 
the initial state contains enough entanglement for the protocols. 
In these latter situations, the lower bounds play a more important role 
than the ranges themselves. 

Our problem is thus to find the maximum and the minimum values of the 
known entanglement measures compatible with a given value of the interference visibility 
subject to the condition that the initial atomic state is the W-like state $|\psi\rangle$ 
given by \eqref{3state} with $c_{1}\ge c_{2}\ge c_{3}>0$.
To focus on genuine tripartite entanglement, the condition $c_{3}>0$ is reminded.  
When the maximum and minimum do not exist, 
the supremum and infimum are calculated, respectively.

In the following, we study the relations between the interference visibility 
and (a) the mixedness of the subsystem, (b) the geometric measure, 
(c) the largest bipartite negativity, and (d) the three-$\pi$. 
The definitions of these entanglement measures 
are summarized in the Appendix. 
These particular measures are chosen to obtain analytical results 
but other entanglement measures could have been studied along the same lines as well. 
Because these calculations are rather tedious, only the final results are shown and 
details will be published elsewhere. 

%%%%%%%%%%%%%%%%%%%
%\subsubsection{Relation to purity}
%%%%%%%%%%%%%%%%%%
\subsubsection{Analytical results} 
%\noindent
(a) \textbf{Mixedness of subsytem}: 
The mixedness of subsystem $M$ of the W-like state $|\psi\rangle$ is given by
\begin{equation}
M=\frac{8}{3}(c_{1}^{2}c_{2}^{2}+c_{2}^{2}c_{3}^{2}+c_{3}^{2}c_{1}^{2})\, .
\end{equation}

When $c_{1}\le c_{2}+c_{3}$, $\vu =1$ and 
the range of mixedness is $2/3\le M\le 8/9$, where the upper bound is attained with the W-state 
and the lower bound is attained by states with $c_{1}=c_{2}+c_{3}$. 

When $c_{1}> c_{2}+c_{3}$ ($\vu<1$), the range is 
\begin{equation}
\frac{2}{3}\left[1-4\frac{\upsilon (1+\upsilon)}{(3+\upsilon)^{2}} \right] \le M<\frac{1+\vu^{2}}{3}\, , 
\end{equation}
where $\upsilon=\sqrt{1-\vu^{2}}$ is the maximal complementary quantity to the visibility. 
The upper bound is obtained for $c_{3}\to 0$ and this case is excluded. 
The lower bound is obtained for $c_{2}=c_{3}$. 
Interestingly, these bounds converge to the same value 2/3 
when one takes the limit $\vu\to 1$ from below. 

%%%%%%%%%%%%%%%%%%
%\subsubsection{Relation to the geometric measure}
%\noindent
(b) \textbf{Geometric measure}: 
The geometric measure $E_{g}$ for three qubits has been studied for the above W-like states 
(\ref{3state}) and has been obtained analytically as a function of the coefficients 
$c_{j}$ \cite{TPT}. If $c_{1}^{2}\le c_{2}^{2}+c_{3}^{2}$ (which implies $c_{1} \le \sqrt{2}/2$),
then $c_{1} \le c_{2}+c_{3}$ and one can construct an acute triangle 
with the coefficients $c_{j}$ being the lengths of its edges. In this case
\begin{align}
E_{g}&=1-4R^{2}\, ,\\
R&=\frac{c_{1}c_{2}c_{3}}{4\sqrt{c_0(c_0-c_{1})(c_0-c_{2})(c_0-c_{3})}}\, ,
\end{align}
where $R$ is the circumradius of the triangle formed by the $c_{j}$ and 
$c_0=(c_{1}+c_{2}+c_{3})/2$. In the other case [i.e., $c_{1}^{2}> c_{2}^{2}+c_{3}^{2}$ 
($c_{1} > \sqrt{2}/2$)] one finds 
\begin{equation}
E_{g}=1-c_{1}^{2}\, .
\end{equation}

When $c_{1}\le c_{2}+c_{3}$, the range of the geometric measure 
is obtained as $1/3\le E_{g}\le 5/9$, where the upper bound is found for the W-state and the lower bound 
is found for states with $c_{1}=\sqrt{2/3}$. 

On the other hand, when $c_{1}> c_{2}+c_{3}$, the geometric measure is bounded by 
\begin{equation}
\frac{1-\upsilon}{3+\upsilon}\le E_{g}<\frac{1-\upsilon}{2}\, ,  
\end{equation}
where the upper bound is obtained for states with $c_{3}\to 0$ 
and the lower bound is obtained for states where $c_{2}=c_{3}$. 
These upper and lower bounds asymptotically reach the values 
$1/2$ and $1/3$, respectively, as $\vu\to 1$ from below.

%%%%%%%%%%%%%%%%%%
%\subsubsection{Relation to bipartite negativity}
%\noindent
(c) \textbf{Largest bipartite negativity}: 
The bipartite negativity of the W-like states with respect to the $j$th qubit is 
given by ${\cal N}_{j}=2c_{j}\sqrt{1-c_{j}^{2}}$. The maximum value achieved 
over all possible bipartite partitions is of interest and thus we consider the quantity 
\begin{equation}
{\cal N}_{{\rm max}} =\max\left\{ {\cal N}_{1},{\cal N}_{2},{\cal N}_{3} \right\}\, . 
\end{equation}

When $c_{1}\le c_{2}+c_{3}$, the range of the maximum bipartite negativity 
is $2\sqrt{2}/3 \le {\cal N}_{{\rm max}} \le 1$. The lower bound is found for states with 
$c_{1}=\sqrt{2/3}$ or $c_{1}=\sqrt{1/3}$ while the upper bound is found 
for the W-state. 

When $c_{1}> c_{2}+c_{3}$, the largest bipartite negativity is bounded by 
\begin{equation}
\sqrt{1-\left(\frac{1+3\upsilon}{3+\upsilon}\right)^{2}}\le {\cal N} _{{\rm max}}<\vu\, ,  
\end{equation}
where the upper bound is found for states with $c_{3}\to0$ 
and the lower bound is found for states with $c_{2}=c_{3}$. 
The upper bound reaches 1 and 
the lower bound converges to $\sqrt{8}/3$ $(=0.9428\dots)$ 
as $\vu$ approaches 1 from below. 

\begin{widetext}
\begin{figure*}[htp]
\centering
\subfigure[]{
\label{fig2a}
\includegraphics[width=6cm]{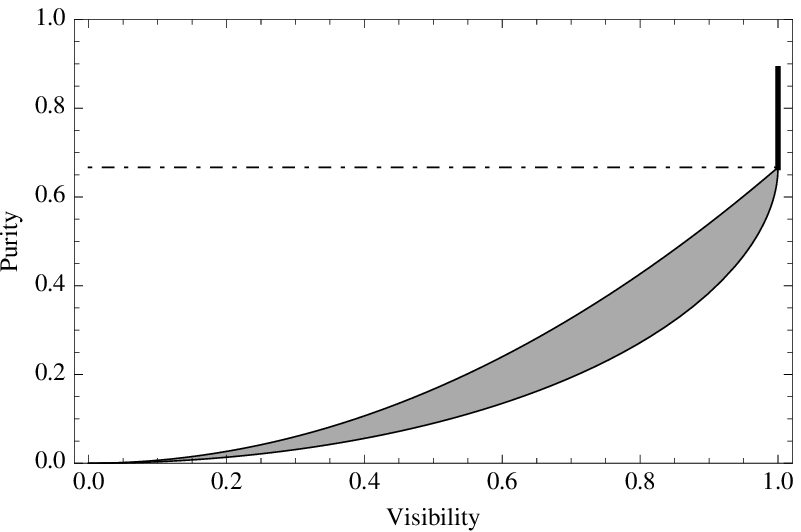}}
\hspace{-0.3cm}
\subfigure[]{
\label{fig2b}
\includegraphics[width=6cm]{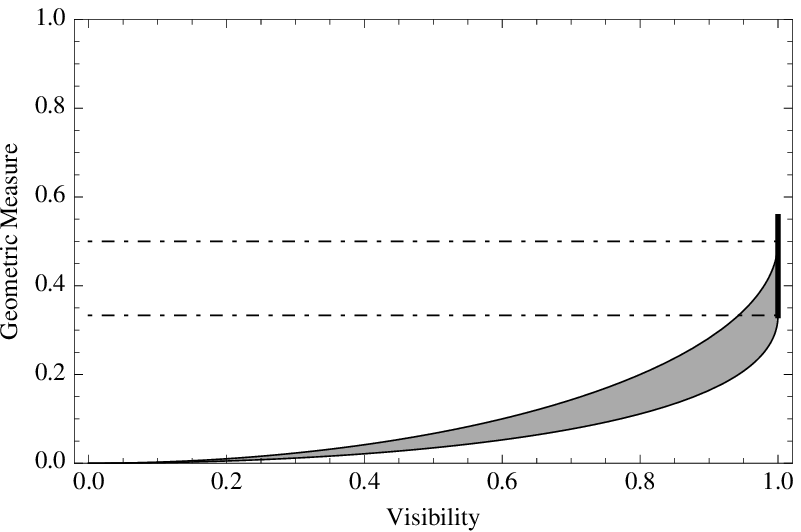}}\\[-0.2cm]
\subfigure[]{
\label{fig2c}
\includegraphics[width=6cm]{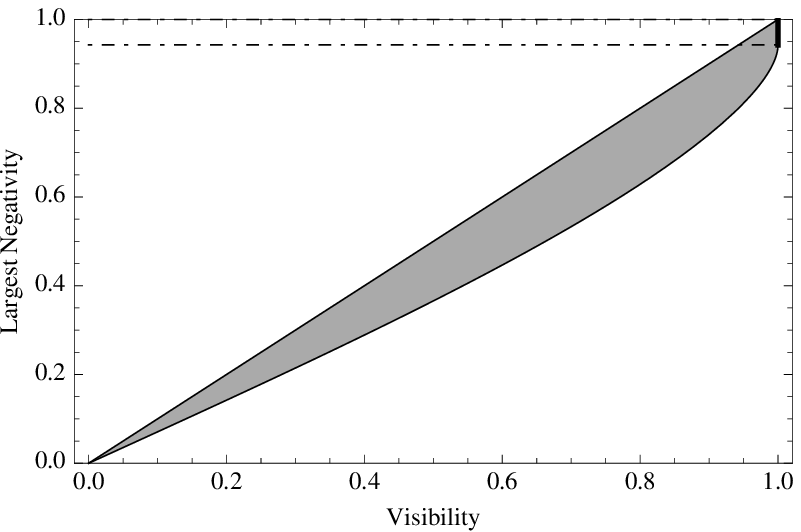}}
\hspace{-0.3cm}
\subfigure[]{
\label{fig2d}
\includegraphics[width=6cm]{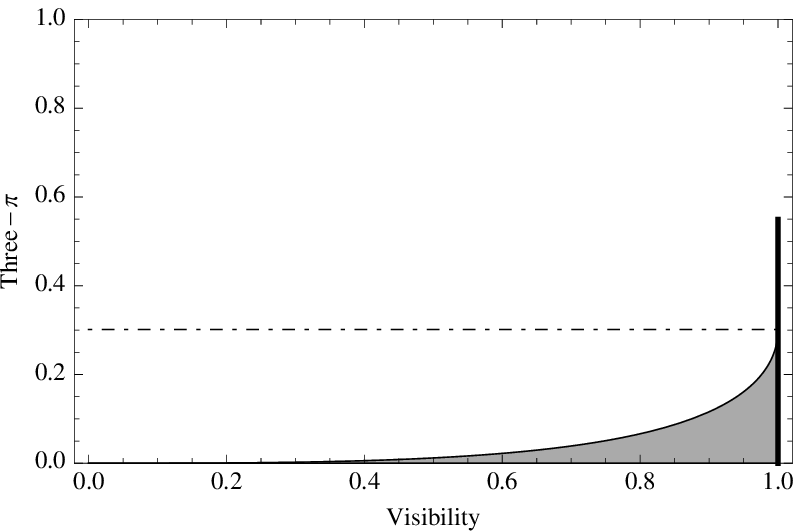}}
\caption{The ranges of entanglement measures studied in this paper as a function of the visibility 
of the far-field interference pattern. The atoms are pinned 
at the vertices of an equilateral triangle, and their mutual distance is large.  
(a) Mixedness of subsystem, (b) geometric measure, 
(c) bipartite negativity, and (d) three-$\pi$, respectively. 
In each graph, the range for $\vu<1$ is shown by two curves filled with a gray area, 
and the thick vertical line indicates the range at $\vu=1$. 
The asymptotic values for $\vu\to 1$ from below are shown by the dashed-dotted lines.}
\end{figure*}
\end{widetext}

%%%%%%%%%%%%%%%%%%
%\subsubsection{Relation to $\pi$-negativity}
%\noindent
(d) \textbf{Three-$\pi$}: 
All entanglement measures studied previously do not vanish when $c_{3}\to 0$. In 
other words, they are insensitive to genuine tripartite entanglement, but are rather 
related to entanglement stemming both from bipartite and tripartite entanglements. 
This is seen from their definitions that all three entanglement measures do not 
vanish when the problem reduces to the two-atom case (i.e., $c_{1}\ge c_{2}>c_{3}=0$). 

To examine the relation to genuine tripartite entanglement measure, 
the so-called three-$\pi$ is examined, which has a similar 
origin as the tangle \cite{CKW,OF}. This is expressed in terms of the $c_{j}$ as
\begin{equation}
{\cal N}_{\pi}=\frac{4}{3}\sum_{j=1}^{3}c_{j}^{4}
\left(\sqrt{1+4c_{1}^{2}c_{2}^{2}c_{3}^{2}/c_{j}^{6}}-1 \right)\, .
\end{equation}
As is easily checked, ${\cal N}_{\pi}=0$ holds when 
$c_{3}\to0$ showing that this entanglement measure is only sensitive 
to genuine tripartite entanglement. 

When $c_{1}\le c_{2}+c_{3}$ the range of the three-$\pi$ is 
$0< {\cal N}_{\pi} \le 4(\sqrt{5}-1)/9$, 
where the upper bound is attained by the W-state 
and the infimum is obtained for states with $c_{3}\to 0$. 
When $c_{1}> c_{2}+c_{3}$, the three-$\pi$ is bounded by 
\begin{equation}
0< {\cal N}_{\pi}\le {\cal N}_{\pi}^{\mathrm{max}} \, ,  
\end{equation}
where the infimum is again obtained for states with $c_{3}\to0$. 
The upper bound is found for the state with 
$c_{2}=c_{3}=\sqrt{(1-\upsilon)/2(3+\upsilon)}$ and reads
\begin{multline}
{\cal N}_{\pi}^{\mathrm{max}}=\frac{2}{3(3+\upsilon)^{2}}[4(1+\upsilon)\sqrt{5+6\upsilon+5\upsilon^{2}}\\
+(1-\upsilon)\sqrt{(1-\upsilon)(17+5\upsilon)}-(9+14\upsilon+9\upsilon^{2})]\, .  
\end{multline}
The upper bound converges to the value 
$2(4\sqrt{5}+\sqrt{17}-9)/27$ $(=0.3012\dots)$ as $\vu\to 1$ from below. 

The results are plotted in Fig.~2, 
(a) mixedness of subsystem, (b) geometric measure, 
(c) bipartite negativity, and (d) three-$\pi$, respectively. 
In each graph, the range for $\vu<1$ is shown by two curves filled with a gray area, 
and the thick vertical line indicates the range at $\vu=1$. 
The asymptotic values for $\vu\to 1$ from below are shown by the dashed-dotted lines.

%%%%%%%%%%%%%%%%%%%%%%%%%%%%%%%%
\subsubsection{Summary of the results}

Let us summarize the obtained relations between the fringe visibility and 
the four entanglement measures and discuss some consequences of them. 

First, the mixedness of the subsystem and the geometric measure 
exhibit similar behaviors showing jumps in the ranges when 
$\vu\to1$ from below. 
These jumps also exist in the three-$\pi$ case, and result 
in detecting the amount of entanglement around 
the full visibility with less accuracy. The appearance of these jumps is partly due to the fact that 
interference visibility defined by \eqref{defvis} is 
a quantity closely related to a bipartite system 
whereas the mixedness of the subsystem and 
geometric measure are related to both bipartite and tripartite entanglement, 
and the three-$\pi$ is only related to tripartite entanglement. 

Second, Fig.~\ref{fig2c} shows that interference visibility 
can reasonably be used to detect the amount of bipartite entanglement 
in the W-like state for all values of interference visibility. 
Remarkably, the upper bound is exactly equal to the largest amount 
of bipartite negativity with respect to all possible bipartite separations 
and does not show any jumps when the visibility approaches its maximal 
value $\vu=1$ from below. 

Last, Fig.~\ref{fig2d} clearly indicates that the interference visibility 
is a poor witness of the genuine tripartite entanglement initially present 
in the atomic system. In particular, getting a far-field interference pattern 
with full visibility $\vu =1$ does not tell anything at all about 
the least amount of tripartite entanglement initially present. 
This result is not really surprising since \eqref{3plim} shows that 
the interference pattern directly depends on the products $c_{i}c_{j}$. 
As mentioned, these products identify with the concurrence of the two-atom 
reduced states obtained by tracing out one atom. It is thus obvious that 
the interference pattern perfectly encodes some bipartite entanglement properties 
of the atomic system, but is less sensitive to genuine tripartite entanglement properties. 

%%%%%%%%%%%%%%%%%%%%%%%%%%%%%%%%
\subsection{State tomography for the W-like states}

In this last section, a state tomography is briefly discussed for 
the W-like states (\ref{3state}) from information contained 
in the far-field interference pattern by observing spontaneously emitted photons. 
The coefficients $c_j$ can be extracted simply by taking a few points of the interference pattern \eqref{3plim}. 
Because of the normalization condition of the state, one needs at least three different points to reconstruct the state. 

When the W-like initial state has complex coefficients $c_{j}\rightarrow c_{j}\exp\I\phi_{j}$, 
the full state tomography also needs 
to find the phases $\phi_{j}$. 
Since the global phase of the state is irrelevant $c_{1}$ can set to be real. 
The remaining phases $\phi_{2}$ and $\phi_{3}$ are 
incorporated in the interference pattern through 
the replacements $\theta_{j}\to\hat{\theta}_{j}$; 
$\hat{\theta}_{1}=\theta_{1}+\phi_{2}-\phi_{3}$, 
$\hat{\theta}_{2} = \theta_{2} - \phi_{3}$, 
$\hat{\theta}_{3} = \theta_{3} + \phi_{2}$. This shows that 
these phases $\phi_{j}$ are merely responsible for a shift of the interference pattern. 
Knowing the position of the fringes, one can extract $\phi_{2}$ 
and $\phi_{3}$, thus completing the state tomography. 
We comment that a more complicated analysis is 
needed when the initial atomic state is mixed. 
The full tomography may well be possible when the atoms are 
separated by a distance on the order of optical atomic transition wavelength.

%%%%%%%%%%%%%%%%%%%%%%%%%%%%%%%%%%%%
%%%%%%%%%%%%%%%%%%%%%%%%%%%%%%%%%%%%
\section{Conclusion and outlook}

In this paper, a system of two-level atoms has been considered when 
initially prepared in the subspace spanned by the first excited atomic states without photons 
and releasing a spontaneous photon 
in the course of time. Some detailed analysis has been given about what kind of 
information the observation of the far-field interference pattern 
can help reveal on the amount of entanglement 
present in the initial atomic state.

In the two-atom case, a previous known result is confirmed showing 
that the interference visibility indeed converges to the concurrence 
of the initial atomic state provided the two atomic qubits are far apart. 
For finite relative distances between the qubits, the deviation of the visibility 
from the concurrence is calculated numerically. 
It is found that this deviation is small whenever the qubits are 
separated by a distance $r$ which is a multiple integer of half the wavelength 
$\lambda_0$ associated to the optical atomic transition. 
A complete state tomography about the state is shown to be possible 
provided the two atoms are sufficiently close to each other. 

In the three-atom case, the family of W-like states has been considered 
and the relations between the interference visibility 
and several known entanglement measures have been analyzed. 
Bounds have been derived 
analytically for the amount of entanglement present in the initial state 
when the interference visibility is given. Our results show 
that the interference visibility is closely related to the bipartite entanglement 
properties of the atoms. 
In particular, it is shown that the interference visibility is directly related 
to the maximal bipartite negativity.

Two possible extensions of this work are to be mentioned. 
First, the possibility of detecting genuine multipartite 
entanglement by defining a suitable interference visibility for multipath 
interferometers. Second, the analysis of the wave-particle duality 
aspects in the three-atom case when the atoms have an internal structure 
allowing for the storage of which-path information. 
These issues shall be investigated in the future.

%%%%%%%%%%%%%%%%%%%%%%%%%%%%%%%%%%%%%
\begin{acknowledgments}

We would like to thank Berge Englert, Dominique Delande, Beno\^{\i}t Gr\'emaud, 
and Cord M\"{u}ller for valuable discussions and useful comments. 
JS would like to thank Berge Englert for his kind hospitality at the Centre for Quantum Technologies 
in Singapore where this work was initiated. JS would also like to thank Dominique Delande 
for his kind hospitality at Laboratoire Kastler Brossel in France where part of this work was done. 
JS and KN would like to acknowledge support by NICT and MEXT. 
ChM is supported by the CNRS PICS 4159 (France), 
by the France-Singapore Merlion program (SpinCold 2.02.07), 
and is partially supported by the National Research Foundation and the Ministry of Education, Singapore.
\end{acknowledgments}

%%%%%%%%%%%%%%%%%%%%%%%%%%%%%%%%%%%%
%%%%%%%%%%%%%%%%%%%%%%%%%%%%%%%%%%%%
\appendix
%%%%%%%%%%%%%%%%%%%%%%%%%%%%%%%%%%%%
\section{Entanglement measures}

Definitions of the entanglement measures used in this paper are listed. 
All these measures take on values between 0 and 1 as a convention 
and are defined for two-qubit and three-qubit systems living in the Hilbert space 
$({\mathbb C}^{2})^{\otimes n}$ ($n=2,3$) of the forms \eqref{2state} 
and \eqref{3state}. For more precise definitions of these entanglement measures and 
for an account of their properties, we refer to Refs.~\cite{PV,Horodecki} and references therein. 

\subsection{Concurrence}
The concurrence of a general two-qubit state $\rho$ is defined by 
\begin{equation}
C(\rho)\equiv \max\{0,\sqrt{\lambda_{1}}-\sqrt{\lambda_{2}}-\sqrt{\lambda_{3}}-\sqrt{\lambda_{4}} \}\, , 
\end{equation}
where $\lambda_{1}\ge\lambda_{2}\ge\lambda_{3}\ge\lambda_{4}$ are 
the eigenvalues of $\sqrt{\rho}\,(\sigma_{y}^{(1)}\!\otimes\sigma^{(2)}_{y})\rho^{*}\,
(\sigma^{(1)}_{y}\!\otimes\sigma^{(2)}_{y})\sqrt{\rho}$. The $y$ component of the Pauli 
spin operator is written as $\sigma^{(j)}_{y}=-\I |e\rangle\langle g|+\I |g\rangle\langle e|$, 
and $\rho^{*}$ denotes the complex conjugate of the state $\rho$. 

\subsection{Negativity}
The negativity of a bipartite system, described by a state $\rho$, 
is defined as twice the sum of the negative eigenvalues of the state $\rho^{T_{j}}$ 
obtained by partial transpose with respect to one subsystem. This can be formally expressed as  
\begin{equation}
{\cal N}(\rho)=|| \rho^{T_{1}}||-1=|| \rho^{T_{2}}||-1\, ,
\end{equation}
where the symbol $T_{j}$ denotes the partial transpose with respect to the $j$th subsystem and 
$||X||$ is the trace class norm of the operator $X$. By convention 
the negativity of the two-qubit Bell states is set to 1. 

For multipartite system, the negativity is defined by the bipartite partitions of 
the system. For example, ${\cal N}_{j}(\rho)=|| \rho^{T_{j}}||-1$ can be defined 
to quantify the entanglement between the $j$th qubit and the rest of the system. 

\subsection{Mixedness of subsystem}
The mixedness of the subsystem as an entanglement measure in a pure state 
$\rho=|\psi\rangle\langle\psi|$ is defined by 
the arithmetical mean of the mixedness associated to each one-qubit reduced states 
obtained from $\rho$. For the three-atom case for example, it reads 
\begin{align} \label{defM}
M(\rho)&=\frac{1}{3}(S(\rho_{1})+S(\rho_{2})+S(\rho_{3}))\, ,\\
S(\rho_j)&=2 (1-\tr \rho_j^{2})\, ,
\end{align}
where $\rho_{1}=\tr_{2,3}\rho$, and so on.
In several papers, this entanglement measure coincides with the definition of visibility 
(e.g., Ref.~\cite{Durr,JB03,Tessier,JB07}). 

\subsection{Geometric measure}

The geometric measure for an $n$-qubit system described by a pure state 
$|\psi\rangle$ is defined by the distance between $|\psi\rangle$ and the closest product state 
\begin{equation}\label{defEg}
E_{g}(\rho)=\min_{|\phi\rangle\in{\rm Pro}}| |\psi\rangle-|\phi\rangle|^{2}\, , 
\end{equation}
where ${\rm Pro}$ represents the space of product states within the Hilbert space 
${\mathbb C}^{\otimes n}$. 

\subsection{Three-$\pi$} 
Several authors have introduced and studied this measure by following the idea 
which led to the tangle in Ref.~\cite{CKW}. The most convenient construction appeared 
in Ref.~\cite{OF} with the definition 
\begin{align} \label{defNpi}
{\cal N}_{\pi}(\rho)&=\frac{1}{3}(\pi_{1}+\pi_{2}+\pi_{3})\, ,\\
\pi_{j}&={\cal N}^{2}_{j}(\rho)+{\cal N}^{2}(\rho_{j})-\sum_{j=1}^{3}{\cal N}^{2}(\rho_{j})\, ,
\end{align}
and ${\cal N}(\rho_{j})$ is the negativity of the reduced state over the $j$th qubit 
[]i.e., $\rho_{j}=\tr_{j}(\rho)$. This can be written more compactly as
\begin{equation}
{\cal N}_{\pi}(\rho)=\frac{1}{3}\sum_{j=1}^{3}[{\cal N}^{2}_{j}(\rho)-2{\cal N}^{2}(\rho_{j})]\, . 
\end{equation}

%%%%%%%%%%%%%%%%%%%%%%%%%%%%%%%%%%%%%
%% References


\begin{thebibliography}{99}
\bibitem{comment2}
This makes perfect sense as the only way to get a separable state 
from $\rho$ is indeed to set $|\rho_{01}|=0$, 
in which case $\rho$ reduces to a weighted statistical mixture 
of the separable states $|0\rangle\langle 0|$ and $|1\rangle\langle 1|$ 
for which no interference is expected in the far-field. In other words, 
coherence has been lost. 
The argument has to be taken with caution as will be seen in Sec.~III B. 
The interference exists in the far-field even for $|\rho_{01}|=0$ if the atoms are close enough.
\end{thebibliography}
\end{document}